\documentclass{article}
\usepackage{ijcai23}

\usepackage{times}
\usepackage{soul}
\usepackage{url}
\usepackage[hidelinks]{hyperref}
\usepackage{multirow}
\usepackage[utf8]{inputenc}
\usepackage[small]{caption}
\usepackage{graphicx}
\usepackage{subcaption}
\usepackage{tikz}
\usepackage{amssymb}
\usepackage{multirow,tabularx}
\usepackage{dirtytalk}
\usepackage{amsmath}
\graphicspath{ {./images/} }
\usepackage{url}
\usepackage{float}
\usepackage{booktabs}
\usepackage{algorithm}
\usepackage{multirow}
\usepackage{algorithmic}
\usepackage[switch]{lineno}
\DeclareUnicodeCharacter{2212}{-}

\title{Measuring Perceived Trust in XAI-Assisted Decision-Making by Eliciting a Mental Model}

\author{
Mohsen Abbaspour Onari$^{1,2}$
\and
Isel Grau$^{1,2}$\and
Marco S. Nobile$^{3}$\And
Yingqian Zhang$^1$
\affiliations
$^1$ Information Systems Group, Eindhoven University of Technology, The Netherlands\\
$^2$ Eindhoven Artificial Intelligence Systems Institute, The Netherlands\\
$^3$ Environmental Sciences, Informatics, and Statistics, Ca' Foscari University of Venice, Italy\\
\emails
\{m.abbaspour.onari, i.d.c.grau.garcia, yqzhang\}@tue.nl,
marco.nobile@unive.it
}

\begin{document}

\maketitle

\begin{abstract}
This empirical study proposes a novel methodology to measure users' perceived trust in an Explainable Artificial Intelligence (XAI) model. To do so, users' mental models are elicited using Fuzzy Cognitive Maps (FCMs). First, we exploit an interpretable Machine Learning (ML) model to classify suspected COVID-19 patients into positive or negative cases. Then, Medical Experts' (MEs) conduct a diagnostic decision-making task based on their knowledge and then prediction and interpretations provided by the XAI model. In order to evaluate the impact of interpretations on perceived trust, explanation satisfaction attributes are rated by MEs through a survey. Then, they are considered as FCM's concepts to determine their influences on each other and, ultimately, on the perceived trust. Moreover, to consider MEs' mental subjectivity, fuzzy linguistic variables are used to determine the strength of influences. After reaching the steady state of FCMs, a quantified value is obtained to measure the perceived trust of each ME. The results show that the quantified values can determine whether MEs trust or distrust the XAI model. We analyze this behavior by comparing the quantified values with MEs' performance in completing diagnostic tasks.
\end{abstract}

\section{Introduction}
\label{sec:intro}

Adopting AI technology into human beliefs and intentions is a difficult task. Especially given that the nature of the task can influence people's attitudes toward algorithmic technology \cite{shin2022user}. To obtain users' trust in an AI agent, they must be confident in its ability to accomplish their goals, particularly in the presence of uncertainty \cite{shin2022user}. Even though advanced Machine Learning (ML) models may accurately predict outcomes, they cannot explain their decision-making process clearly. This weakness caused arising of Explainable Artificial Intelligence (XAI) field to tackle the need for interpretability in ML models. As a result, XAI aims to enable the practical use of ML models in critical domains by enhancing their interpretability \cite{wysocki2023assessing}. The general agreement is that making ML algorithms transparent is a means of establishing user trust \cite{bitzer2023algorithmic}. Despite widespread interdisciplinary attention to trust in AI, a widely accepted definition and a clear understanding of how it relates to other constructs, such as user understanding or satisfaction, is required. Additionally, measuring trust in AI and the relationship between different measures are still open research questions \cite{benk2022value}.

This empirical research proposes a methodology to measure and quantify medical experts' (MEs) perceived trust in an interpretable ML model by eliciting their mental models, including subjectivity. Our method consists of three phases. First, we will exploit an interpretable ML model with high predictive performance to classify suspected COVID-19 patients. Second, MEs will conduct a diagnostic task. We want to study whether the interpretable ML model can positively contribute to solving the diagnostic task by MEs. Third, MEs will indicate their satisfaction with the model's interpretation by rating explanation satisfaction criteria to solve the task through a survey. MEs will also determine the influence of explanation satisfaction criteria on each other and how they can establish their perceived trust in the model. After completing the survey, we can elicit MEs' trust mental model using Fuzzy Cognitive Maps (FCMs) method. FCM facilitates the enhanced integration of expert, stakeholder, and indigenous knowledge by creating scenarios that connect quantitative analysis and qualitative storylines \cite{jetter2014fuzzy}. Utilizing FCM to represent mental models can foster an understanding of how individuals filter, process, and store information. At the same time, it can elucidate how these perspectives influence decision-making and actions within a specific context \cite{gray2013fuzzy}. It has been proven that structuring human knowledge through collecting FCMs offers useful applications beyond the mere characterization of conventional expert systems. Moreover, it provides a means to represent community understanding through scaled-up mental modeling \cite{gray2013fuzzy}. To involve MEs' subjectivity, fuzzy linguistic variables are used to develop FCMs, and the Mean of Maxima (MoM) method will be used for defuzzification.

The contributions of this paper can be summarized as follows:
\begin{itemize}
  \item Proposing an empirical experiment based on a decision-making task to study the impact of an interpretable ML model on solving it.
  \item Designing a survey to evaluate the MEs' satisfaction with the model's interpretation and its impact on establishing trust.
  \item Eliciting MEs' trust mental model by FCMs based on the designed survey and quantifying perceived trust.
  \item Considering MEs' mental subjectivity using fuzzy linguistic variables to develop FCMs.
\end{itemize}

The remainder of the present research is organized as follows. Section \ref{sec:literature} reviews the literature on trust in the XAI field. Section \ref{sec:preliminaries} presents the fundamental concepts of this research. Section \ref{sec:case-study} covers the proposed methodology to measure perceived trust based on a real-world case study. Section \ref{results} provides the results of the proposed methodology. Finally, Section \ref{conclusion} presents the conclusion and directions for future studies.

\section{Related work}
\label{sec:literature}

Trust in an AI model is eroded when users cannot comprehend the reasons behind observed actions or decisions \cite{miller2019explanation}. There are ongoing research questions regarding measuring trust in AI and the relationship between various measures. \cite{hoffman2018metrics} presented the first comprehensive study to scale trust and user satisfaction on XAI. \cite{schmidt2019quantifying} proposed a quantitative measure for the quality of interpretable methods and users' trust in ML decisions. \cite{lakkaraju2020fool} established a user study with domain experts to show how user trust in black box models can be manipulated via misleading explanations. \cite{yang2020visual} explored how providing example-based explanations for an ML classifier affects the appropriate trust of end-users. In a virtual reality setting for public transportation users, \cite{faulhaber2021effect} demonstrated that trust propensity could predict explicit trust. This study evaluated explicit trust through a questionnaire and implicit trust by analyzing the participant's behavior. To evaluate combined explanation methods for reinforcement learning, \cite{huber2021local} conducted a user study to examine participants' mental models to investigate whether their trust was appropriate given the agents' capabilities. Users' mental models were elicited through a retrospection task, and their satisfaction was evaluated through a survey. \cite{shin2022user} investigated the factors that explain users' trust and intention to use the primary XAI fake news detector through a survey. \cite{dikmen2022effects} analyzed the significance of domain expertise during an interaction with an XAI system in a financial investment context. The findings demonstrated that participants equipped with domain knowledge were less dependent on the AI assistant when it provided incorrect information. In an exploratory experiment, \cite{leichtmann2023effects} assessed the effects of XAI methods and educational intervention on AI-assisted decision-making behavior. Besides, they asked participants to rate how much they trusted the AI model and comprehended the classification of the AI on a 5-point Likert scale. \cite{ueno2023trust} explored the roles of trust and reliance as critical elements of the XAI user experience. They measured trust on a 7-point Likert scale of 12 items: 7 for trust and 5 for distrust.

Analyzing existing research in the literature demonstrates the high significance of qualitative research, i.e., questionnaire and survey, in evaluating trust. One of the limitations of most studies in the literature is the need for more structured methodologies to elicit users' mental models. Qualitative research must also be able to elicit users' mental models to show how their trust is established. We will achieve this objective in the current study by developing an FCM model for each ME.

\section{Preliminaries}
\label{sec:preliminaries}

\subsection{Definition of trust}
\label{sec:trust}

When making critical decisions, dealing with incomplete, conflicting, and uncertain information is possible, as it is often impossible to access perfect knowledge in many areas. This uncertainty creates a situation where the decision-maker risks making incorrect decisions and trusting the wrong entity \cite{cho2015survey}. Given that trust is a multidisciplinary concept, \cite{cho2015survey} proposed a common definition of trust across different disciplines as follows:

\textit{``The willingness of the trustor (evaluator) to take a risk by relying on their subjective belief that the trustee (evaluatee) will act reliably to maximize their interests, particularly under uncertainty due to conflicting or absent evidence. This belief is based on the trustor's cognitive assessment of past experiences with the trustee.''}

The concept of trustworthiness is seen in AI fields frequently. However, it should be stressed that trust and trustworthiness are entirely disentangled concepts, in which pursuing one does not entail pursuing the other \cite{jacovi2021formalizing}. When a user trusts the AI model, anticipation depends on whether the model can carry out its contract. So \cite{jacovi2021formalizing} declare that an AI model is trustworthy to some contract if it can maintain this contract. Now, reaching a more comprehensive definition of trust in the AI domain is possible. Here, we refer to the definition of \cite {jacovi2021formalizing} for Human-AI trust:

\textit{``If H (human) perceives that M (AI model) is trustworthy to contract C, and accepts vulnerability to M’s actions, then H trusts M contractually to C. The objective of H in trusting M is to anticipate that M will maintain C in the presence of uncertainty; consequently, trust does not exist if H does not perceive risk.''}

\subsection{XAI and perceived trust}
\label{sec:XAI-trust}

AI models require explainability to justify their decisions to various stakeholders. The justification is crucial in building trust among users, so it is necessary to clarify the AI model's decisions \cite{kaur2022trustworthy}. XAI field tries to bring algorithmic transparency by developing post-hoc and inherently interpretable models. \cite{miller2022we} states that trust as a mental attitude must be measured in field studies, lab experiments, and surveys/interviews with human participants. These works have centered on precisely defining the components of trust to enable the measurement of perceived trust. It is crucial to measure perceived trust as it impacts the acceptance, adoption, and dependence on a particular agent. However, asking participants to state their trust is not equivalent to demonstrating trust, which involves deciding whether to use an agent for their actions \cite{miller2022we}. While the measurement of perceived trust has received significant attention in the literature, the existing techniques do not measure the impact of transparency, interpretability, and explainability methods on participants' trust \cite{miller2022we}. In this research, we will use the mentioned concepts to study whether they help solve tasks and elicit users' trust mental model.

\subsection{Eliciting mental models and measuring perceived trust by FCMs}
\label{sec:fcm}

Kosko introduced FCMs \cite{kosko1986fuzzy} to mitigate the limited ability of cognitive maps \cite{axelrod1976cognitive} in representing causal beliefs in social scientific knowledge \cite{napoles2014two}. Multiple domain experts who possess knowledge in a particular area contribute as knowledge engineers to manually develop an FCM or a mental model \cite{papageorgiou2011new}. They start by identifying the key domain components or concepts ($C$) and then determine the influence (edges) of concepts on each other ($w$). Finally, they create a graph that displays the concepts, edges, and fuzzy strengths of edges, representing the achieved FCM \cite{papageorgiou2011new}. The values of $C$ and $w$ are usually considered in the interval [0, 1] and [− 1, 1], respectively. There are three types of relationships between concepts in the FCM \cite{onari2021risk}:

\begin{itemize}
  \item $w_{ij}>0$, direct influence between the concepts of $C_{i}$ and $C_{j}$,
  \item $w_{ij}<0$, inverse influence between the concepts of $C_{i}$ and $C_{j}$,
  \item $w_{ij}=0$, no relationship between the concepts of $C_{i}$ and $C_{j}$.
\end{itemize}
For the reasoning process of an FCM, the following simple mathematical formula is used:

\begin{flalign}
\label{fcm-formula}
    &A^{(k+1)}_{i}=f\left(\sum_{j=1,j\neq i}^{N}A^{(k)}_{j}.w_{ji} \right) &&
\end{flalign}
which in Eq. \ref{fcm-formula}, $A^{(k)}_{i}$ is the state vector representing values of $C_{i}$ at the iteration $k$. $f(x)$ is an activation function that is usually either sigmoid or hyperbolic tangent to keep the states vector's values between [0, 1] and [-1, 1], respectively.

\begin{figure}[!ht]
    \centering
        \includegraphics[width=0.5\textwidth]{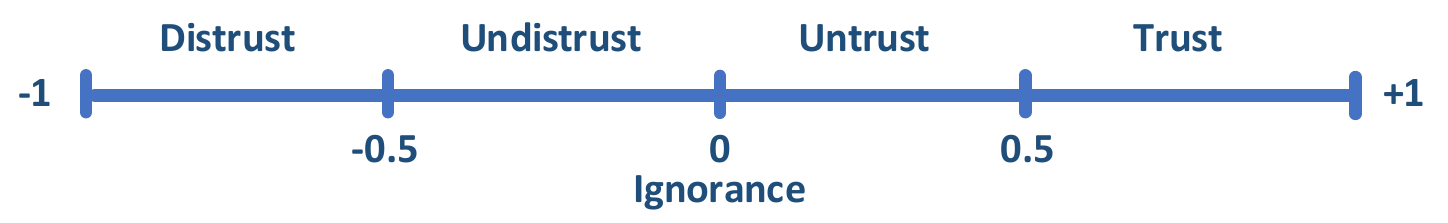}
    \caption{Trust continuum used to quantify the trust level of MEs.}
    \label{fig:trust-scale}
\end{figure}

To elicit MEs' mental models to measure perceived trust, we follow the study of \cite{hoffman2018metrics}. In this study, they developed an explanation satisfaction scale that users rate their understandability, satisfaction with the explanations' quality, and trust in the AI system. \cite{miller2022we} declares that the presented trust scale by Hoffman does not explicitly measure the effect of trust because it is not a process. In practice, any type of trust is tentative, and trust in machines undergoes a dynamic evolution as individuals interact with the machine over time. Moreover, trust is not a binary concept but rather a continuum that can be more accurately represented by probability or confidence \cite{miller2022we}. We want to add another shortcoming of Hoffman's proposed trust scale as missing the subjectivity of the users, which has been discussed in \cite{cho2015survey}. To overcome the first shortcoming, we want to propose two solutions. First, we design an experiment to involve MEs in a diagnostic task with and without the assistance of an XAI model. The objective is to make an interaction between MEs and the XAI model to study whether it positively contributes to solving the task. Second, we use the trust continuum proposed by \cite{cho2015survey} to quantify the trust level of MEs in the XAI model (See Fig. \ref{fig:trust-scale}). This figure shows that the trust continuum is a value in the interval [-1, 1]. So if the target concept's value in FCM is less than -0.5, the ME distrust the XAI model, and values higher than 0.5 show their trust in it. Also, the value of 0 indicates the ignorance value. To obtain the same scale by FCM, we use hyperbolic tangent as an activation function.

In the developed mental model of MEs, we use fuzzy linguistic variables to rate explanation satisfaction attributes and their strength of influence on each other. This approach can address the second shortcoming. \cite{cho2015survey} declared that including the subjectivity of trust is crucial and can be studied by fuzzy set theory \cite{zadeh1983role} with a high capability in modeling uncertainty. To develop FCMs, it is possible to defuzzify fuzzy numbers to crisp numbers by MoM method. After implementing FCMs and reaching stability, we will have a quantified value to measure the perceived trust.

\section{Case study}
\label{sec:case-study}

\subsection{COVID-19 data set}
\label{sec:covid-19}

In the early advent of COVID-19, it was crucial to rapidly determine suspected patients' status upon arrival at the emergency department. To achieve this, a blood sample is routinely taken and tested for at least 30 distinct measurements. These measurements are analyzed to identify any subtle changes that may indicate the presence of COVID-19. The data set consists of 12873 patients with 32 clinical values derived from blood samples utilized as features to train ML models \cite{onari2022comparing}.

The data set was collected between July 14, 2019, and July 20, 2020. The authors followed ethical aspects of AI application by signing written agreements regarding the limited use of the data. Second, they adhered to security measures to protect data privacy per the agreements. Third, patients' identities were anonymized. The data set includes missing values in both the features and labels. Certain observations were collected before the COVID-19 outbreak and were classified as negative cases. Observations with no labels and missing values exceeding 40\% were discarded as they offer no meaningful information for the ML model. Patients under the age of 18 were also excluded. Ultimately, the data set comprises 8781 observations, of which 8461 are negative and 320 are positive.

\subsection{Phase 1: Exploiting interpretable ML model to classify patients}
\label{sec:phase1}

In \cite{onari2022comparing}, we showed the high predictive performance of Repeated Incremental Pruning to Produce Error Reduction (RIPPER) \cite{cohen1995fast} on the introduced data set. RIPPER is an interpretable ML algorithm that operates on rules directly learned from the data. It applies a depth-first search and generates one rule at a time through rule induction. RIPPER produces IF-THEN classification rules using the separate-and-conquer technique and the reduced error pruning approach. Afterward, a set of rules is returned, which can be applied to classify new objects \cite{onari2022comparing}. 

Before implementing RIPPER, to handle missing values in the data set, K-Nearest Neighbor (KNN) data imputation is applied to fill out missing values. Then, the correlation between the features is calculated and features with a high correlation with each other are dropped from the data set, leaving 27 features to build the ML model. The data set is split into training and test data sets with 80\%-20\% partition, respectively. Even though RIPPER shows a high capability on imbalanced data set classification, we implement SMOTE oversampling technique to have the same number of positive and negative cases in the training phase. Furthermore, the model's hyperparameters are optimized using the Grid Search technique. The model undergoes 5-fold cross-validation on the training data set to validate its performance. The results are then reported on the test data set. Table \ref{table:results} shows the predictive performance of RIPPER based on four different metrics.

\begin{table}[!ht]
    \centering
    \caption{RIPPER’s predictive performance}
    \label{table:results}
    \begin{tabular}{lcccc}
        \cline{2-5}
         & Accuracy & Precision & Recall & F1 score \\
        \hline
        RIPPER & 0.9841 & 0.8667 & 0.6393 & 0.7358\\
        \hline
    \end{tabular}
\end{table}

The RIPPER model obtained 98.41\% accuracy in the correct prediction of patients and 73.58\% F1 score, which is considered good performance for this classification problem \cite{onari2022comparing}. To interpret the prediction's logic, RIPPER generates three rules on the test data set represented in Tables \ref{table:rule1}-\ref{table:rule3}. It should be mentioned that RIPPER considers the group with the most number of patients (in this case, negative cases) as the default group and makes rules by classifying groups with less number of patients (in this case, positive cases). After generating rules for positive cases, it considers the remaining instances negative.

\begin{table*}[!ht]
    \centering
    \caption{Rule 1 generated by RIPPER to classify patients into positive cases}
    \label{table:rule1}
    \begin{tabular}{lcccccccc}
        \cline{2-7}
               & Albumin & Alkaline phosphatase & Calcium & Erythrocytes & Glucose & Lactate dehydrogenase\\
        \hline
        Rule 1 & $\le37.9$ (AND)& $\le82$ (AND)& $\le2.28$ (AND)& $\ge3.94$ (AND)& $\ge5.66$ (AND)& $\ge302$\\
        \hline
    \end{tabular}
\end{table*}

\begin{table*}[!ht]
    \centering
    \caption{Rule 2 generated by RIPPER to classify patients into positive cases}
    \label{table:rule2}
    \begin{tabular}{lccccc}
        \cline{2-6}
               & Alkaline phosphatase & Basophils & C-reactive Protein & Leukocytes & Lipase\\
        \hline
        Rule 2 & $\le83.6$ (AND)& $\le0.01$ (AND)& $\ge19.62$ (AND)& $\le7.69$ (AND)& $\ge30.5$\\
        \hline
    \end{tabular}
\end{table*}

\begin{table*}[!ht]
    \centering
    \caption{Rule 3 generated by RIPPER to classify patients into positive cases}
    \label{table:rule3}
    \begin{tabular}{lcccc}
        \cline{2-5}
               & Erythrocytes & Lactate dehydrogenase & Leukocytes & Mean Cellular Haemoglobin\\
        \hline
        Rule 3 & $\ge4.29$ (AND)& $\ge320$ (AND)& $\le7.68$ (AND)& $\ge1.85$\\
        \hline
    \end{tabular}
\end{table*}

\subsection{Phase 2: Diagnostic task of MEs}
\label{sec:phase2}

MEs conduct a decision-making task to evaluate the functionality and efficiency of RIPPER's prediction and interpretation to diagnose the disease. The diagnostic task is conducted in two independent sub-tasks: first, based on MEs' knowledge, and then RIPPER's prediction and rules. Our main assumption is that using the interpretable model can positively contribute to solving the task by MEs. Hence, four patients are selected from the test data set, providing only features used by RIPPER's rules. The main reason for providing fewer features than the original ones is to consider the presence of uncertainty in the trust measurement. So, by using the features used by RIPPER, we rely on its rationale in diagnosing the disease. Table \ref{table:patients} demonstrates four selected patients' blood sample test results. In this table, the ground truth (GT) represents the recorded status of the patients at the hospital, followed by RIPPER's prediction. We selected two instances in which GT status and RIPPER's prediction have the same results and two with contradictory results in GT and RIPPER's prediction.

\begin{table*}[!ht]
    \centering
    \caption{Selected patients' blood sample test results}
    \label{table:patients}
    \begin{tabular}{llcccc}
        \cline{2-6}
               & Features & Patient 1 & Patient 2 & Patient 3 & Patient 4\\
        \hline
        1 & Albumin & 37 & 44.2 & 36.7 & 38.6 \\
        2 & Alkaline phosphatase & 70 & 82 & 69 & 70.2\\
        3 & Basophils & 0.03 & 0.02 & 0.01 & 0.03\\
        4 & Calcium & 2.09 & 2.5 & 2.18 & 2.29\\
        5 & C-reactive protein & 1.43 & 4.68 & 39.91 & 173\\
        6 & Erythrocytes & 4.75 & 4.66 & 4.87 & 4.34\\
        8 & Glucose & 10.36 & 6.53 & 5.4 & 6.87\\
        9 & Lactate dehydrogenase & 392 & 280 & 481 & 154\\
        10 & Leukocytes & 7.13 & 4.5 & 6.53 & 16.6\\
        11 & Lipase & 47.8 & 16.2 & 66.2 & 29.1\\
        12 & Mean Cellular Haemoglobin & 1.895 & 1.931 & 1.848 & 1.751\\
        \hline
         & GT & Positive COVID-19 & Negative COVID-19 & Negative COVID-19 & Positive COVID-19\\
         & RIPPER prediction & Positive COVID-19 & Negative COVID-19 & Positive COVID-19 & Negative COVID-19\\
        \hline
    \end{tabular}
\end{table*}

\subsection{Phase 3: Eliciting MEs' mental model by FCM}
\label{sec:phase3}

We elicit MEs' mental models and measure perceived trust in three phases.

\textbf{Step 1: Explanation satisfaction attributes as FCM's concepts.}

After finishing the diagnostic task phase by MEs, their satisfaction with the quality of the provided rules to solve the task is evaluated. You can find explanation satisfaction attributes in Table \ref{table:satisfaction-attributes}. To consider the subjectivity of the experts, instead of using the Likert scale to rate explanation satisfaction attributes, we use five fuzzy linguistic variables represented in Table \ref{table:fuzzy-concept}. The explanation satisfaction attributes are then considered as FCM's concepts, and their corresponding defuzzified values are their activation value in implementing FCM.

\begin{table*}[!ht]
    \centering
    \caption{Explanation satisfactions attributes}
    \label{table:satisfaction-attributes}
    \begin{tabular}{lll}
        \hline
        Concept & Key attributes of explanation & Explanation\\
        \hline
        C1 & Understandability & The rules were understandable to diagnose the disease. \\
        C2 & Sufficiency of details & The rules had sufficient details to help me to diagnose the disease. \\
        C3 & Completeness & The rules were complete enough to diagnose the disease. \\
        C4 & Feeling of satisfaction & I am satisfied with the quality of the rules for diagnosing the disease. \\
        C5 & Accuracy & The rules were accurate enough to diagnose the disease.\\
        C6 & Usability & The rules are easy to use for diagnosing the disease. \\
        C7 & Functionality & In general, the rules helped me in my task of diagnosing the disease. \\
        \hline
    \end{tabular}
\end{table*}

\begin{table}[!ht]
    \centering
    \caption{Linguistic variables to rate explanation satisfaction criteria}
    \label{table:fuzzy-concept}
    \resizebox{\columnwidth}{!}{%
    \begin{tabular}{llcc}
        \cline{2-4}
         & Linguistic variables & Membership function & Defuzzified value\\
        \hline
        1 & I disagree strongly & (0, 0, 0.25) & 0\\
        2 & I disagree somewhat & (0, 0.25, 0.5) & 0.25\\
        3 & I’m neutral about it & (0.25, 0.5, 0.75) & 0.5\\
        4 & I agree somewhat & (0.5, 0.75, 1) & 0.75\\
        5 & I agree strongly & (0.75, 1, 1) & 1\\
        \hline
    \end{tabular}
    }
\end{table}

\textbf{Step 2: Determining the influence of concepts on each other and MEs' perceived trust.}

In this step, MEs determine the influence of concepts on each other with their strength. This step creates a big picture of their mental model to construct their perceived trust in the XAI model. To do so, an 8×8 matrix is built, embedding concepts and one additional target concept: trust. MEs using fuzzy linguistic variables represented in Table \ref{table:fuzzy-weights} determine the strength of influence.

\begin{table}[!ht]
    \centering
    \caption{Linguistic variables to determine the strength of influence}
    \label{table:fuzzy-weights}
    \resizebox{\columnwidth}{!}{%
    \begin{tabular}{llcc}
        \cline{2-4}
         & Linguistic variables & Membership function & Defuzzified value\\
        \hline
        1 & Inversely high & (-1, -1, -0.5) & -1\\
        2 & Inversely low & (-1, -0.5, 0) & -0.5\\
        3 & No influence & (-0.5, 0, 0.5) & 0\\
        4 & Directly low & (0, 0.5, 1) & 0.5\\
        5 & Directly high & (0.5, 1, 1) & 1\\
        \hline
    \end{tabular}
    }
\end{table}

There are three types of influence: inverse, nothing, and direct, and two types of strength of influence: low and high. The following examples illustrate the meaning of these fuzzy linguistic variables:

\begin{itemize}
  \item Inversely high: with an increment in the sufficiency of details, I obtain a strong decrement in the degree of usability.
  \item Inversely low: with an increment in the sufficiency of details, I obtain a small decrement in the degree of usability.
  \item No influence: A change in the sufficiency of details does not influence the feeling of satisfaction.
  \item Directly low: with an increment in the sufficiency of details, I obtain a small increment in understandability.
  \item Directly high: with an increment in the sufficiency of details, I obtain a strong increment in understandability.
\end{itemize}

In addition to determining the influence of concepts on each other, the influence of concepts on the MEs' trust (target concept) is also determined here. Consequently, each ME will have a unique mental model.

\textbf{Step 3: Implementing FCM to quantify perceived trust.}

FCMs are implemented based on Eq. \ref{fcm-formula}, and interaction between concepts of FCMs continues until occurring one of the following states \cite{beena2011structural}:

\begin{itemize}
  \item A stable state is reached.
  \item A limit cycle is reached.
  \item Chaotic behavior is exhibited.
\end{itemize}

After reaching stability, the quantified value of the target concept is used to measure perceived trust for each ME.

\section{Results}
\label{results}

\begin{table*}[!ht]
    \centering
    \caption{The result of the diagnostic task by MEs}
    \label{table:diagnostic-task}
        \resizebox{\textwidth}{!}{%
        \begin{tabular}{cccccccccc}
        \cline{3-4} \cline{5-6} \cline{7-8} \cline{9-10}
        \multirow{2}{*}{} & \multirow{2}{*}{} & \multicolumn{2}{c}{ME1} & \multicolumn{2}{c}{ME2} & \multicolumn{2}{c}{ME3} & \multicolumn{2}{c}{ME4} \\
        \cline{3-4} \cline{5-6} \cline{7-8} \cline{9-10}
            & & Knowledge & XAI-Assistant & Knowledge & XAI-Assistant & Knowledge & XAI-Assistant & Knowledge & XAI-Assistant\\ \hline

        \multirow{2}{*}{Patient 1} & GT = Positive & $\times$ & & IP &  & \checkmark & & $\times$ &\\ 
        \cline{3-4} \cline{5-6} \cline{7-8} \cline{9-10}
                                   & XAI = Positive &  & \checkmark & & IP & & \checkmark &  & IP\\ \hline

        \multirow{2}{*}{Patient 2} & GT = Negative & \checkmark &  & IP & & $\times$ & & $\times$ &\\ 
        \cline{3-4} \cline{5-6} \cline{7-8} \cline{9-10}
                                   & XAI = Negative &  & \checkmark &  & IP && \checkmark & & $\times$\\ \hline

        \multirow{2}{*}{Patient 3} & GT = Negative & $\times$ &  & $\times$ &  & \checkmark & & $\times$ & \\ 
        \cline{3-4} \cline{5-6} \cline{7-8} \cline{9-10}
                                   & XAI = Positive &  & \checkmark &  & IP & & $\times$ & & \checkmark\\ \hline

        \multirow{2}{*}{Patient 4} & GT = Positive & IP &  & IP & & \checkmark & & \checkmark & \\ 
        \cline{3-4} \cline{5-6} \cline{7-8} \cline{9-10}
                                   & XAI = Negative &  & \checkmark & & IP & & $\times$ & & \checkmark\\ \hline
        \end{tabular}%
        }
\end{table*}

\begin{table*}[!ht]
    \centering
    \caption{Fuzzy membership function to express the level of satisfaction with rules}
    \label{table:me-concepts}
    \begin{tabular}{ccccc}
    \hline
        Concept ~ & ME1 & ME2 & ME3 & ME4 \\ \hline
        C1 & I agree somewhat & I agree strongly & I agree strongly & I’m neutral about it \\ \hline
        C2 & I agree somewhat & I’m neutral about it & I agree somewhat & I agree somewhat \\ \hline
        C3 & I disagree somewhat & I disagree strongly & I agree somewhat & I’m neutral about it \\ \hline
        C4 & I’m neutral about it & I agree somewhat & I agree strongly & I disagree somewhat \\ \hline
        C5 & I disagree strongly & I disagree somewhat & I agree somewhat & I disagree somewhat\\ \hline
        C6 & I agree somewhat & I agree strongly & I agree somewhat & I’m neutral about it \\ \hline
        C7  & I’m neutral about it  & I’m neutral about it  & I agree strongly  & I’m neutral about it \\ \hline
    \end{tabular}
\end{table*}

\begin{figure*}[!htb]
    % subfigure 1
    \begin{subfigure}{0.49\textwidth}
	\includegraphics[width=\textwidth]{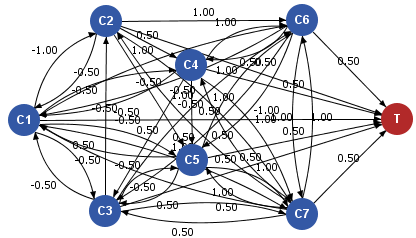}
	\caption{ME1}
	\end{subfigure}
	% subfigure 2
	\begin{subfigure}{0.49\textwidth}
	\includegraphics[width=\textwidth]{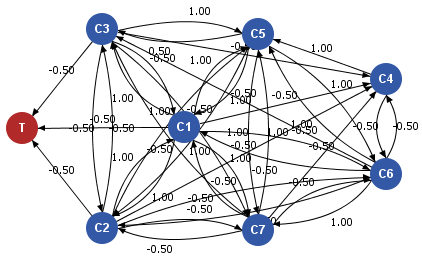}
	\caption{ME2}
	\end{subfigure}
 
	% subfigure 3
	\begin{subfigure}{0.49\textwidth}
	\includegraphics[width=\textwidth]{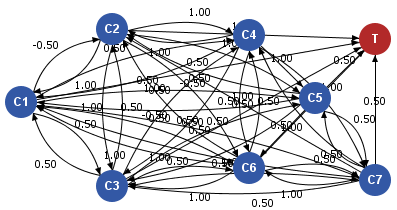}
	\caption{ME3} 
	\end{subfigure}
	% subfigure 4
	\begin{subfigure}{0.41\textwidth}
	\includegraphics[width=\textwidth]{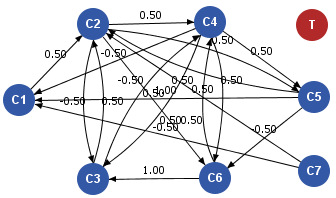}
	\caption{ME4}
	\end{subfigure}
	
	\captionsetup{justification=justified}
	\caption{Mental models (FCMs) of perceived trust for each ME. The concept representing the attribute Trust is highlighted in red.}
\label{fig:mental-models}
\end{figure*}

This study has been monitored and approved by the Ethical Board of the university. We invited survey participants via email and provided them with an informed consent form and an attached information sheet about the study. We stored and analyzed data within the secured premises of the university. We have collected the answers from four MEs but expect to receive more responses. 

The MEs are general practitioners with at least two years of working experience. They all have working experience in healthcare centers during the COVID-19 pandemic. The survey starts by providing general information and confirming the consent to participate in the study. Then, the participants provide general information, including expertise and working experience background. After that, the diagnostic task is initiated. The results of the diagnostic task are shown in Table \ref{table:diagnostic-task}. The check mark symbol shows a consensus between MEs' diagnosis with GT and XAI prediction. The multiply symbol indicates that the MEs' diagnosis contradicts GT and XAI prediction. Finally, IP shows that MEs believe diagnosing the disease is impossible with the provided clinical tests and XAI model.

The results show that ME1 misdiagnosed Patient 1 as negative based on their knowledge. However, in the second sub-task, they relied on XAI and could diagnose the case correctly. For Patient 2, in both sub-tasks, ME1 diagnosed the case as negative. For Patient 3, ME1 is sure that the case has positive COVID-19 and has the same opinion in both sub-tasks. In the last case, ME1 believes it is impossible to diagnose the case based on the provided test results. However, in the second sub-task, ME1 relied on the XAI model. We can conclude that in two cases, ME1 trusts in the XAI model to solve the tasks, showing it has a positive contribution. 

On the other hand, ME2 did not trust the XAI model in all cases and believes it is impossible to diagnose the cases based on it. ME2 has the same idea about the provided blood sample test results as well. ME2 solved only one case (Patient 3), which misdiagnosed it. 

ME3 only has one misdiagnosis based on their knowledge (Patient 2). However, in the second sub-task, ME3 relied on the XAI to solve the task correctly. ME3 entirely relied on their knowledge for the remaining cases, and even in cases where XAI had wrong predictions, ME3 diagnosed the cases correctly. We can conclude ME3 mostly relies on their knowledge but also can use XAI efficiently. 

Finally, although ME4 mostly misdiagnosed cases, they preferred to rely on their knowledge. ME4, in only one case (Patient 4), relied on the XAI model, resulting in misdiagnosing the case that already diagnosed it correctly in the first sub-task. We can conclude that ME4 is not optimistic about the XAI model but somehow uses it as well.

After completing the diagnostic task, MEs' satisfaction with the model's interpretation is rated. According to Table \ref{table:me-concepts}, ME3 has highly satisfied with the XAI model. On the other side, ME4 is mostly neutral about the provided rules and has neither feelings of satisfaction nor dissatisfaction with the XAI model. For the remaining experts, their satisfaction does not inform a particular pattern.

Figure \ref{fig:mental-models} shows the mental models for each ME, built using the FCMExpert tool \cite{napoles2017fuzzy,napoles2018fcm}. A quick look at this figure shows that the mental model of ME1 consists of positive and negative influences of concepts on each other. From our point of view, ME1 has the most challenging mental model as it has no specific prediction pattern. Also, satisfaction with the model's interpretation of ME1 has the same pattern. It should be reminded that we concluded in the diagnostic task that the XAI model positively contributed to solving two cases by ME1.

The mental model of ME2 has many inverse influences between concepts. Besides, ME2, except for three concepts that inversely influence their trust, is neutral about the influence of concepts on trust. In the diagnosis task, ME2 did not find the XAI model reliable in diagnosing the disease. It seems that ME2 distrusts the model, and we expect to obtain a high value of distrust. 

On the other hand, ME3 is optimistic about the XAI model because they could use it efficiently in the diagnostic task and also has considered many direct influences between concepts. Besides, ME3, in general, was also satisfied with the XAI model. Except for one concept, all concepts directly influence their trust, and we can expect a high trust value. 

Finally, ME4 could not find any relation between concepts and their trust. It turns out that based on the trust continuum, we expect the value of 0 for trust, which shows ME4 has a feeling of ignorance about the XAI model. We observed the same behavior in the diagnostic task and satisfaction with the model's interpretations. 

Now that the FCM models are ready, we can run an inference process to compute the trust value for each ME. Table \ref{table:trust-valaue} shows the result of FCM to quantify trust.

\begin{table}[!ht]
    \centering
    \caption{Quantified value of trust}
    \label{table:trust-valaue}
    \begin{tabular}{ccccc}
    \cline{2-5}
        ~ & ME1 & ME2 & ME3 & ME4 \\ \hline
        Trust value & 0.8079 & -0.7645 & 0.9992 & 0\\ \hline
    \end{tabular}
\end{table}

Based on FCM, we see that the trust value of ME1 is 0.8079, which is a considerable amount. It shows that, in general, ME1 trusts the model. ME2 distrusted the XAI model because FCM obtained a value of -0.7645, less than the -0.5 threshold value. Also, ME3 has obtained a high trust value close to 1, meaning they trust the XAI model. As expected, the trust value of ME4 is 0, showing that ME4 completely ignores the XAI model and neither trust nor distrust it. 

\section{Conclusion and discussion}
\label{conclusion}

In this empirical study, a methodology was proposed to measure and quantify the perceived trust of MEs in an XAI model by eliciting their mental model. The value of trust is quantified by developing and implementing an FCM for each ME. We analyzed MEs' mental models based on their performance in the diagnostic tasks and developed FCMs. We can say that we could track the trust behavior of MEs in both scenarios. The performance of the MEs in the diagnostic tasks overlaps with the obtained perceived trust value by FCM.

This work is still in the first development steps and has some limitations we will address in future studies. First, we need to involve more MEs to have a statistically reliable conclusion about the efficiency of our proposed methodology. Second, triangular fuzzy linguistic variables were used in this study to model MEs' subjectivity and develop FCM simultaneously. It is crucial to model subjectivity with other fuzzy sets to calibrate the trust value and study the effect of fuzzy sets type in the ultimate quantified trust value. Third, we will consider aggregating all MEs' mental models to have a global trust quantified value. It can help us study the impact of generalizing mental models on the final trust value. Finally, we will consider presenting different types of explanations to MEs to study their impact on improving their working experience with them and possibly enhance their trust in the XAI model.

\newpage

%% The file named.bst is a bibliography style file for BibTeX 0.99c
\bibliographystyle{named}
\bibliography{ijcai23}

\begin{thebibliography}{}

\bibitem[\protect\citeauthoryear{Axelrod}{1976}]{axelrod1976cognitive}
Robert Axelrod.
\newblock The cognitive maps of political elites, 1976.

\bibitem[\protect\citeauthoryear{Beena and Ganguli}{2011}]{beena2011structural}
P~Beena and Ranjan Ganguli.
\newblock Structural damage detection using fuzzy cognitive maps and hebbian
  learning.
\newblock {\em Applied soft computing}, 11(1):1014--1020, 2011.

\bibitem[\protect\citeauthoryear{Benk \bgroup \em et al.\egroup
  }{2022}]{benk2022value}
Michaela Benk, Suzanne Tolmeijer, Florian von Wangenheim, and Andrea Ferrario.
\newblock The value of measuring trust in ai-a socio-technical system
  perspective.
\newblock {\em arXiv preprint arXiv:2204.13480}, 2022.

\bibitem[\protect\citeauthoryear{Bitzer \bgroup \em et al.\egroup
  }{2023}]{bitzer2023algorithmic}
Tobias Bitzer, Martin Wiener, and W~Alec Cram.
\newblock Algorithmic transparency: Concepts, antecedents, and consequences--a
  review and research framework.
\newblock {\em Communications of the Association for Information Systems},
  52(1):16, 2023.

\bibitem[\protect\citeauthoryear{Cho \bgroup \em et al.\egroup
  }{2015}]{cho2015survey}
Jin-Hee Cho, Kevin Chan, and Sibel Adali.
\newblock A survey on trust modeling.
\newblock {\em ACM Computing Surveys (CSUR)}, 48(2):1--40, 2015.

\bibitem[\protect\citeauthoryear{Cohen}{1995}]{cohen1995fast}
William~W Cohen.
\newblock Fast effective rule induction.
\newblock In {\em Machine learning proceedings 1995}, pages 115--123. Elsevier,
  1995.

\bibitem[\protect\citeauthoryear{Dikmen and Burns}{2022}]{dikmen2022effects}
Murat Dikmen and Catherine Burns.
\newblock The effects of domain knowledge on trust in explainable ai and task
  performance: A case of peer-to-peer lending.
\newblock {\em International Journal of Human-Computer Studies}, 162:102792,
  2022.

\bibitem[\protect\citeauthoryear{Faulhaber \bgroup \em et al.\egroup
  }{2021}]{faulhaber2021effect}
Anja~K Faulhaber, Ina Ni, and Ludger Schmidt.
\newblock The effect of explanations on trust in an assistance system for
  public transport users and the role of the propensity to trust.
\newblock In {\em Proceedings of Mensch und Computer 2021}, pages 303--310.
  2021.

\bibitem[\protect\citeauthoryear{Gray \bgroup \em et al.\egroup
  }{2013}]{gray2013fuzzy}
Steven~A Gray, Erin Zanre, and Stefan~RJ Gray.
\newblock Fuzzy cognitive maps as representations of mental models and group
  beliefs.
\newblock In {\em Fuzzy cognitive maps for applied sciences and engineering:
  From fundamentals to extensions and learning algorithms}, pages 29--48.
  Springer, 2013.

\bibitem[\protect\citeauthoryear{Hoffman \bgroup \em et al.\egroup
  }{2018}]{hoffman2018metrics}
Robert~R Hoffman, Shane~T Mueller, Gary Klein, and Jordan Litman.
\newblock Metrics for explainable ai: Challenges and prospects.
\newblock {\em arXiv preprint arXiv:1812.04608}, 2018.

\bibitem[\protect\citeauthoryear{Huber \bgroup \em et al.\egroup
  }{2021}]{huber2021local}
Tobias Huber, Katharina Weitz, Elisabeth Andr{\'e}, and Ofra Amir.
\newblock Local and global explanations of agent behavior: Integrating strategy
  summaries with saliency maps.
\newblock {\em Artificial Intelligence}, 301:103571, 2021.

\bibitem[\protect\citeauthoryear{Jacovi \bgroup \em et al.\egroup
  }{2021}]{jacovi2021formalizing}
Alon Jacovi, Ana Marasovi{\'c}, Tim Miller, and Yoav Goldberg.
\newblock Formalizing trust in artificial intelligence: Prerequisites, causes
  and goals of human trust in ai.
\newblock In {\em Proceedings of the 2021 ACM conference on fairness,
  accountability, and transparency}, pages 624--635, 2021.

\bibitem[\protect\citeauthoryear{Jetter and Kok}{2014}]{jetter2014fuzzy}
Antonie~J Jetter and Kasper Kok.
\newblock Fuzzy cognitive maps for futures studies—a methodological
  assessment of concepts and methods.
\newblock {\em Futures}, 61:45--57, 2014.

\bibitem[\protect\citeauthoryear{Kaur \bgroup \em et al.\egroup
  }{2022}]{kaur2022trustworthy}
Davinder Kaur, Suleyman Uslu, Kaley~J Rittichier, and Arjan Durresi.
\newblock Trustworthy artificial intelligence: a review.
\newblock {\em ACM Computing Surveys (CSUR)}, 55(2):1--38, 2022.

\bibitem[\protect\citeauthoryear{Kosko}{1986}]{kosko1986fuzzy}
Bart Kosko.
\newblock Fuzzy cognitive maps.
\newblock {\em International journal of man-machine studies}, 24(1):65--75,
  1986.

\bibitem[\protect\citeauthoryear{Lakkaraju and
  Bastani}{2020}]{lakkaraju2020fool}
Himabindu Lakkaraju and Osbert Bastani.
\newblock " how do i fool you?" manipulating user trust via misleading black
  box explanations.
\newblock In {\em Proceedings of the AAAI/ACM Conference on AI, Ethics, and
  Society}, pages 79--85, 2020.

\bibitem[\protect\citeauthoryear{Leichtmann \bgroup \em et al.\egroup
  }{2023}]{leichtmann2023effects}
Benedikt Leichtmann, Christina Humer, Andreas Hinterreiter, Marc Streit, and
  Martina Mara.
\newblock Effects of explainable artificial intelligence on trust and human
  behavior in a high-risk decision task.
\newblock {\em Computers in Human Behavior}, 139:107539, 2023.

\bibitem[\protect\citeauthoryear{Miller}{2019}]{miller2019explanation}
Tim Miller.
\newblock Explanation in artificial intelligence: Insights from the social
  sciences.
\newblock {\em Artificial intelligence}, 267:1--38, 2019.

\bibitem[\protect\citeauthoryear{Miller}{2022}]{miller2022we}
Tim Miller.
\newblock Are we measuring trust correctly in explainability, interpretability,
  and transparency research?
\newblock {\em arXiv preprint arXiv:2209.00651}, 2022.

\bibitem[\protect\citeauthoryear{N{\'a}poles \bgroup \em et al.\egroup
  }{2014}]{napoles2014two}
Gonzalo N{\'a}poles, Isel Grau, Rafael Bello, and Ricardo Grau.
\newblock Two-steps learning of fuzzy cognitive maps for prediction and
  knowledge discovery on the hiv-1 drug resistance.
\newblock {\em Expert Systems with Applications}, 41(3):821--830, 2014.

\bibitem[\protect\citeauthoryear{N{\'a}poles \bgroup \em et al.\egroup
  }{2017}]{napoles2017fuzzy}
Gonzalo N{\'a}poles, Maikel Leon, Isel Grau, and Koen Vanhoof.
\newblock Fuzzy cognitive maps tool for scenario analysis and pattern
  classification.
\newblock In {\em 2017 IEEE 29th International Conference on Tools with
  Artificial Intelligence (ICTAI)}, pages 644--651. IEEE, 2017.

\bibitem[\protect\citeauthoryear{N{\'a}poles \bgroup \em et al.\egroup
  }{2018}]{napoles2018fcm}
Gonzalo N{\'a}poles, Maikel~Leon Espinosa, Isel Grau, and Koen Vanhoof.
\newblock {FCM Expert}: software tool for scenario analysis and pattern
  classification based on fuzzy cognitive maps.
\newblock {\em International Journal on Artificial Intelligence Tools},
  27(07):1860010, 2018.

\bibitem[\protect\citeauthoryear{Onari \bgroup \em et al.\egroup
  }{2021}]{onari2021risk}
Mohsen~Abbaspour Onari, Samuel Yousefi, and Mustafa~Jahangoshai Rezaee.
\newblock Risk assessment in discrete production processes considering
  uncertainty and reliability: Z-number multi-stage fuzzy cognitive map with
  fuzzy learning algorithm.
\newblock {\em Artificial Intelligence Review}, 54(2):1349--1383, 2021.

\bibitem[\protect\citeauthoryear{Onari \bgroup \em et al.\egroup
  }{2022}]{onari2022comparing}
Mohsen~Abbaspour Onari, Marco~S Nobile, Isel Grau, Caro Fuchs, Yingqian Zhang,
  Arjen-Kars Boer, and Volkher Scharnhorst.
\newblock Comparing interpretable ai approaches for the clinical environment:
  an application to covid-19.
\newblock In {\em 2022 IEEE Conference on Computational Intelligence in
  Bioinformatics and Computational Biology (CIBCB)}, pages 1--8. IEEE, 2022.

\bibitem[\protect\citeauthoryear{Papageorgiou}{2011}]{papageorgiou2011new}
Elpiniki~I Papageorgiou.
\newblock A new methodology for decisions in medical informatics using fuzzy
  cognitive maps based on fuzzy rule-extraction techniques.
\newblock {\em Applied Soft Computing}, 11(1):500--513, 2011.

\bibitem[\protect\citeauthoryear{Schmidt and
  Biessmann}{2019}]{schmidt2019quantifying}
Philipp Schmidt and Felix Biessmann.
\newblock Quantifying interpretability and trust in machine learning systems.
\newblock {\em arXiv preprint arXiv:1901.08558}, 2019.

\bibitem[\protect\citeauthoryear{Shin and Chan-Olmsted}{2022}]{shin2022user}
Jieun Shin and Sylvia Chan-Olmsted.
\newblock User perceptions and trust of explainable machine learning fake news
  detectors.
\newblock {\em International Journal of Communication}, 17:23, 2022.

\bibitem[\protect\citeauthoryear{Ueno \bgroup \em et al.\egroup
  }{2023}]{ueno2023trust}
Takane Ueno, Yeongdae Kim, Hiroki Oura, and Katie Seaborn.
\newblock Trust and reliance in consensus-based explanations from an
  anti-misinformation agent.
\newblock In {\em Extended Abstracts of the 2023 CHI Conference on Human
  Factors in Computing Systems}, pages 1--7, 2023.

\bibitem[\protect\citeauthoryear{Wysocki \bgroup \em et al.\egroup
  }{2023}]{wysocki2023assessing}
Oskar Wysocki, Jessica~Katharine Davies, Markel Vigo, Anne~Caroline Armstrong,
  D{\'o}nal Landers, Rebecca Lee, and Andr{\'e} Freitas.
\newblock Assessing the communication gap between ai models and healthcare
  professionals: explainability, utility and trust in ai-driven clinical
  decision-making.
\newblock {\em Artificial Intelligence}, 316:103839, 2023.

\bibitem[\protect\citeauthoryear{Yang \bgroup \em et al.\egroup
  }{2020}]{yang2020visual}
Fumeng Yang, Zhuanyi Huang, Jean Scholtz, and Dustin~L Arendt.
\newblock How do visual explanations foster end users' appropriate trust in
  machine learning?
\newblock In {\em Proceedings of the 25th International Conference on
  Intelligent User Interfaces}, pages 189--201, 2020.

\bibitem[\protect\citeauthoryear{Zadeh}{1983}]{zadeh1983role}
Lotfi~Asker Zadeh.
\newblock The role of fuzzy logic in the management of uncertainty in expert
  systems.
\newblock {\em Fuzzy sets and systems}, 11(1-3):199--227, 1983.

\end{thebibliography}

\end{document}